\documentclass[11pt,a4paper]{article}
\usepackage{jheppub}
\usepackage{natbib}
\usepackage{graphicx}
\usepackage{cases}
\usepackage{xcolor}

\usepackage{mathtools}
\usepackage{hyperref}
\usepackage{float}
\usepackage{comment}
\usepackage{multirow}

\def\mnras{MNRAS}

\definecolor{Blu}{rgb}{0.,0.,1.}

\author[a]{Koun Choi,}
\author[b,c]{Injun Jeong,}
\author[b,c]{Sunghyun Kang,}
\author[b,c,d]{Arpan Kar,}
\author[b,c]{Stefano Scopel,}
\affiliation[a]{Institute for Basic Science (IBS), 55 Expo-ro, Yuseong-gu, Daejeon, 34126, South Korea}
\affiliation[b]{Center for Quantum Spacetime, Sogang University, 35 Baekbeom-ro, Mapo-gu, Seoul, 121-742, South Korea}
\affiliation[c]{Department of Physics, Sogang University, 35 Baekbeom-ro, Mapo-gu, Seoul, 121-742, South Korea}
\affiliation[d]{Laboratoire de Physique Théorique et Hautes Énergies (LPTHE), CNRS $\&$ Sorbonne Université, 4 Place Jussieu, F-75252, Paris, France}
\emailAdd{koun@ibs.re.kr}
\emailAdd{natson@naver.com}
\emailAdd{francis735@naver.com}
\emailAdd{arpankarphys@gmail.com}
\emailAdd{scopel@sogang.ac.kr}

\title{Sensitivity of WIMP bounds on the velocity distribution in the limit of a massless mediator}

\abstract{
We discuss the sensitivity of the bounds on the spin--independent (SI) and spin--dependent (SD) WIMP--proton and WIMP--neutron interaction couplings $\alpha_{SI, SD}^{p,n}$ on the WIMP velocity distribution for a massless mediator. We update the bounds in the Standard Halo Model (SHM) for direct detection and the neutrino signal from WIMP annihilation in the Sun (fixing the annihilation channel to $b\bar{b}$), and set a halo--independent bound for the first time using the single--stream method. 
In the case of a massless mediator the SHM capture rate in the Sun diverges and is regularized by removing the contribution of WIMPs locked into orbits that extend beyond the Sun--Jupiter distance. We discuss the dependence of the SHM bounds on the Jupiter cut showing that it can be sizeable for $\alpha_{\rm SD}^p$ and a WIMP mass $m_\chi$ exceeding 1 TeV. Our updated SHM bounds show an improvement between about two and three orders of magnitude compared to the previous ones in the literature. Our halo--independent analysis shows that, with the exception of $\alpha_{\rm SD}^p$ at large $m_\chi$, the relaxation of the bounds compared to the SHM is of the same order of that for contact interactions, i.e. relatively moderate in the low and high WIMP mass regimes and as large as $\sim 10^2$ for $m_\chi\simeq$ 20 GeV. On the other hand, the exact determination of the relaxation of the bound becomes not reliable for $\alpha_{\rm SD}^p$ and  $m_\chi\gtrsim$ 1 TeV due to the sensitivity of the SHM capture rate in the Sun to the details of the Maxwellian velocity distribution at low incoming WIMP speeds. In contrast, the halo--independent bounds are robust against the details of the velocity distribution including the Jupiter cut and the local escape speed, as expected.
} 


\begin{document}
\hspace*{87.0mm}{CQUeST-2024-0739}\\
\maketitle

\section{Introduction}
\label{sec:introduction}

Cold Dark Matter (CDM) is believed to be indispensable to explain the existence of galaxies and clusters of galaxies as we see them today, and to provide about 25\% of the density of the Universe~\cite{Planck:2018vyg}. The most popular CDM candidates are Weakly Interacting Massive Particles (WIMPs), whose phenomenology is characterized by the fact that the same interactions that are assumed to keep WIMPs in thermal equilibrium with Standard Model (SM) particles in the Early Universe and that allow to predict its relic density in agreement to observation through the thermal decoupling mechanism, provide an opportunity to detect them experimentally today. The main process that can be used to search for WIMPs is their scattering process off nuclear targets, that enters at the same time in Direct Detection (DD) experiments, that search for the recoil energy of nuclei in solid--state, 
liquid and gaseous detectors in underground laboratories shielded 
against cosmic rays~\cite{DD_Goodman1984, DD_LEWIN1996, JUNGMAN1996, DD_Schumann2019, Snowmass_Leane2022}, 
or in experiments searching for neutrinos produced by WIMP annihilation inside celestial bodies (Earth, Sun), 
where the WIMPs are accumulated after being captured through 
the same WIMP--nucleus scattering process 
that enters DD~\cite{cap_nu_sun_PhysRevLett1985, cap_nu_sun_HAGELIN1986, dm_cap_nu_SREDNICKI1987, Jungman:1994jr, idm_sun_Catena_2018}.  

An important piece of input in the calculation of expected signals is the WIMP speed distribution $f(u)$ in the reference frame of the Solar system, that determines the WIMP incoming flux\footnote{Neglecting the relative velocity between the Earth and the Sun both present direct detection experiments and signals from WIMP capture in the Sun are sensitive to the speed distribution $f(u) \equiv \int d\Omega f(\vec{u}) u^2$.}. In particular, for the $f(u)$ both early analytical estimations~\cite{violent_relaxation} and more recent numerical models of Galaxy formation~\cite{VDF_Lacroix2020, VDF_Lopes2020} are compatible to the Standard Halo Model (SHM) scenario, where the $f(u)$ is given by a Maxwellian in the galactic halo rest frame~\cite{SHM_1986, SHM_1988}. However, although the SHM provides a useful zero--order approximation to describe the WIMP speed distribution, numerical simulations of Galaxy formation can only shed light on statistical average properties of galactic halos, while our lack of information about the specific merger history of the Milky Way prevents us to rule out the possibility that the $f(u)$ has sizeable non--thermal components. 
Recently, data released by Gaia have opened the possibility to track the merger history of the Milky way, revealing many local stellar streams formed by material stripped from merged satellites. This may suggest the presence of dark matter streams or of a dark disc~\cite{Lisanti:2011as,OHare:2018trr,Necib:2019zbk,Necib:2019zka}. In particular, such components may remain with a small velocity dispersion and co-rotating with the Galactic disc, resulting in a substantial contribution to the  low speed tail of the $f(u)$ in the reference frame of the Solar system.

Based on the above considerations, several attempts have been made to develop halo--independent approaches with the goal to remove the dependence of the experimental bounds on the choice of a specific $f(u)$, both for the WIMP--nucleus and for WIMP--electron scattering process~\cite{halo_independent_2010, halo_independent_Fox_2010, halo_uncertainty_Frandsen2011, astrophysics_independent_Herrero-Garcia2012, halo_independent_DelNobile_2013, Bozorgnia:2013hsa, halo_independent_Fox2014, halo_independent_Feldstein2014, halo_independent_Scopel_inelastic_2014, halo_independent_Feldstein2014_2, halo_independent_Bozorgnia2014, halo_independent_Anderson2015, Halo-independent_Ferrer2015, halo_independent_Kahlhoefer, Gondolo_Scopel_2017, halo_independent_Catena_Ibarra_2018, velocity_uncertainty_Ibarra2018, velocity_independent_2019, halo_independent_electron_scattering, Herrera_2024_information_divergences}, i.e. to work out the most conservative bounds from null searches compatible with the only constraint:

\begin{equation}
    \int_{u=0}^{\infty} f(u) du = 1,
    \label{eq:f_normalization}
\end{equation}

\noindent but allowing for any possible speed profile of the distribution. In order to do so it is necessary to combine DD with capture in the Sun: in fact, due to the experimental threshold, DD searches are not sensitive to small WIMP speeds while capture in the Sun is suppressed for fast WIMPs. Only when combined together the complementarity of the two detection techniques allow to probe the $f(u)$ in the full range of WIMP speeds $u$. 

In the past, using this approach Ref.~\cite{Halo-independent_Ferrer2015} developed a particularly straightforward method that allows to obtain conservative constraints that are independent of the $f(u)$ and only requires the assumption~(\ref{eq:f_normalization}). In the following we will refer to this procedure as the single--stream method. In Ref.~\cite{halo_independent_sogang_2023} it was used to obtain halo--independent bounds in the case of the most general non-relativistic effective Hamiltonian that drives the scattering process off nuclei of a WIMP of spin 1/2, while in~\cite{halo_independent_inelastic_sogang_2023} the single--stream analysis was extended to the case of Inelastic Dark Matter~\cite{inelastic_Tucker-Smith:2001}. In both cases a contact interaction was assumed. In the following we wish to further extend the single--stream method to the case of an interaction with a massless mediator.

In past years, several proposal have been put forward where the DM particle interacts with a gauge boson in the dark sector.  Searches for such mediator have also been pursued in accelerators~\cite{Ilten:2022lfq}, dark matter searches~\cite{Mitridate:2022tnv} or neutrino telescopes~\cite{Batell:2022xau}.  In the following we will be agnostic about the specific particle--physics model describing such WIMP--nucleon interaction, and use the fact that both the DD and capture processes are non--relativistic to parameterize
the WIMP--nucleon interaction with an effective Hamiltonian ${\bf\mathcal{H}}$ that complies with Galilean symmetry~\cite{nreft_haxton1,nreft_haxton2}.  In particular, to zero--th order in the WIMP--nucleon relative velocity $\vec{v}$ and momentum transfer $\vec{q}$ the effective Hamiltonian ${\bf\mathcal{H}}$  consists of the usual spin--independent (SI) and spin--dependent (SD) interaction terms. Specifically, in the general case the effective Hamiltonian is given by:

\begin{equation}
    {\cal H}=\sum_{\tau=0,1}\left ( \frac{\alpha_{\rm SI}^\tau}{q^2+M_0^2}1_\chi 1_N t^\tau + \frac{\alpha_{\rm SD}^\tau}{q^2+M_0^2} \vec{S}_\chi \cdot \vec{S}_N t^\tau\right )
    \label{eq:H_long},
\end{equation}

\noindent with $M_0$ the mass of the mediator. In the following we will study the present sensitivities to the $\alpha_{\rm SI}^\tau$, $\alpha_{\rm SD}^\tau$ couplings in the case when $M_0=0$. Moreover, for completeness we will compare some of our results to those for the corresponding couplings for a contact interaction $c_{\rm SI}^\tau$, $c_{\rm SD}^\tau$, defined as:

\begin{eqnarray}
    &&\lim_{M_0\rightarrow \infty}\frac{\alpha^{\tau}_{\rm SI}}{q^2+M_0^2}\rightarrow 
    \frac{\alpha^{\tau}_{\rm SI}}{M_0^2}\equiv c^{\tau}_{\rm SI}
    \nonumber\\
    &&\lim_{M_0\rightarrow \infty}\frac{\alpha^{\tau}_{\rm SD}}{q^2+M_0^2}\rightarrow 
    \frac{\alpha^{\tau}_{\rm SD}}{M_0^2}\equiv c^{\tau}_{\rm SD}.
\end{eqnarray}

In Eq.~(\ref{eq:H_long}) $1_{\chi}$  and $1_{N}$ are identity operators, $q$ is the transferred momentum, $\vec{S}_{\chi}$ and
$\vec{S}_{N}$ are the WIMP and nucleon spins, respectively, while
 $t^0$ = $1$ , $t^1$ = $\sigma^3$ denote the 2 $\times$ 2 identity and third Pauli matrix in isospin space, respectively, and the isoscalar and isovector coupling constants $\alpha^0_j$ and $\alpha^1_j$ (with $j=SI, SD$) are related to those to protons and neutrons $\alpha^p_j$ and $\alpha^n_j$ by $\alpha^0_j$ = $\alpha^p_j + \alpha^n_j$ and $\alpha^1_j$ = $\alpha^p_j - \alpha^n_j$ (the same holds for contact interactions, $c^0_j$ = $c^p_j + c^n_j$ and $c^1_j$ = $c^p_j - c^n_j$).

The first goal of the present paper is to update the existing bounds~\cite{Liang:2013dsa,OHare:2018trr} on the two long--range couplings $\alpha_{\rm SI}^{p,n}$, $\alpha_{\rm SD}^{p,n}$ from the combinations of capture in the Sun and direct detection in the standard case of a Maxwellian WIMP velocity distribution in the Galaxy. Moreover, we wish to discuss the sensitivity of such bounds on the WIMP velocity distribution by obtaining halo--independent constraints using the single--stream method~\cite{Halo-independent_Ferrer2015}. 

Compared to the case of a contact interaction, the WIMP--nucleon Hamiltonian of Eq.~(\ref{eq:H_long}) can potentially imply: (i) in the SHM case an improved sensitivity of capture in the Sun versus DD; (ii) an enhanced sensitivity of the results on the specific choice of the velocity distribution. The latter aspect can be understood by noticing that Eq.~(\ref{eq:H_long}) leads to a divergent WIMP--nucleus cross section when $q\rightarrow$ 0 (an indication that in such kinematic regime a specific ultraviolet completion of the model is required).  This is not an issue in the case of DD experiments, where a lower cut--off on $q$ is determined by the experimental threshold. However, the kinematic conditions for a WIMP to be captured in the Sun extend to $q\rightarrow$ 0, so that using the Hamiltonian in~(\ref{eq:H_long}) the capture rate in this case is formally infinite. Indeed, when expressed in term of the mediator mass $M_0$, the capture rate diverges logarithmically for $M_0\rightarrow$ 0 when the thermal motion of the nuclear targets in the Sun is neglected~\cite{capture_long_range_Fan_2013, capture_long_range_Chen_2015}, and quadratically when it is taken into account~\cite{capture_long_range_Gaidau_2021}. 
The origin of this divergence arises from the low-velocity tail of the distribution $f(u)$ and is due to the fact that near $u$ = 0 DM particles far from the Sun have a very small kinetic energy. As a result, even scattering interactions yielding very small (actually, {\it arbitrarily} small) recoil energies can result in a negative total energy, i.e. in the dark matter particle being captured. 

The approach used in the literature to address this issue is to count as captured only DM particles confined to orbits that lie within Jupiter’s orbit~\cite{jupiter_cut_Kumar_2012,Liang:2013dsa}. This introduces a lower cut--off on $q$ and yields a finite result for the capture rate. Thanks to the smaller momentum transfers involved, the constraints obtained in this way from capture in the Sun can be potentially more competitive than those from DD~\cite{Liang:2013dsa,OHare:2018trr}. 
However, the sensitivity to the low--speed range of the $f(u)$ of the calculation of capture in the Sun in the case of a massless mediator potentially limits the robustness of such bounds, while the regularizing procedure of the Jupiter bound appears arbitrary. As we will show, a halo--independent approach removes both problems, and this represents one of the main motivations of the present analysis. In the following we will specifically address this issue and, more in general, obtain halo--independent constraints to discuss their sensitivity on the WIMP velocity distribution.

The plan of the paper is the following. In Section~\ref{sec:wimp_nucleon_scattering} we provide the expressions for the expected signals obtained from Eq.~(\ref{eq:H_long}). In particular, in Section~\ref{sec:dd} we summarize those for the expected rate in DD experiments, while in Section~\ref{sec:capture} those for capture in the Sun. In Section~\ref{sec:single_stream} we summarise the single--stream method that we use to obtain our halo--independent bounds. Our quantitative results both for the SHM and in the halo--independent case are contained in Section~\ref{sec:analysis}. Finally, conclusions are drawn in Section~\ref{sec:conclusions}. Appendix~\ref{app:experiments} contains the details of the experimental observations that we employ for DD (LZ~\cite{LZ_2022}, XENON--nT~\cite{xenon_nT}, XENON--1T~\cite{xenon_2018} in Appendix~\ref{app:lz}, PICO--60(C$_3$F$_8$)~\cite{pico60_2019} in Appendix~\ref{app:pico60_c3f8} and PICO--60(CF$_3$I)~\cite{pico60_2015} in Appendix~\ref{app:pico60cf3i}) and those for neutrino telescopes (Super--Kamiokande~\cite{SuperK_2015}, IceCube~\cite{IceCube:2016} and projections for Hyper--Kamiokande \cite{Bell:2021esh} in Appendix~\ref{app:NT}).

\section{WIMP-nucleon scattering in the limit of a massless mediator}
\label{sec:wimp_nucleon_scattering}

For $M_0=0$ the WIMP--nucleus scattering amplitude for the Hamiltonian given in Eq.~\eqref{eq:H_long} is given by~\cite{nreft_haxton1,nreft_haxton2}:
\begin{eqnarray}
&&\frac{1}{2 j_{\chi}+1} \frac{1}{2 j_{T}+1}|\mathcal{M}_T(q^2)|^2 = \nonumber\\
&&=\frac{4\pi}{2 j_{T}+1} \sum_{\tau,\tau^{\prime}=0,1}  
\left \{
\frac{\alpha_{\rm SI}^{\tau}\alpha_{\rm SI}^{\tau^{\prime}}}{q^4}W_{T M}^{\tau\tau^{\prime}}(q)+\frac{j_{\chi}(j_{\chi}+1)}{12}\frac{\alpha_{\rm SD}^{\tau}\alpha_{\rm SD}^{\tau^{\prime}}}{q^4}\left [ W_{T \Sigma^{\prime}}^{\tau\tau^{\prime}}(q)+W_{T \Sigma^{\prime\prime}}^{\tau\tau^{\prime}}(q) \right ]
\right \}
 \,.
\label{eq:squared_amplitude}
\end{eqnarray}
\noindent In the equation above $j_{\chi}$ and $j_{T}$ are the WIMP and the target nucleus spins respectively, while the $W^{\tau\tau^{\prime}}_{T k}$'s (with $k=M,\Sigma^{\prime},\Sigma^{\prime\prime}$ ) are the nuclear response functions (or nuclear form factors) defined in Refs.~\cite{nreft_haxton1,nreft_haxton2,Catena_nuclear_form_factors}. 

The differential cross section for the WIMP--nucleus scattering can be written in terms of the recoil energy ($E_R={q^2}/{2 m_T}$) as:
\begin{equation}
\frac{d\sigma_T}{d E_R}(q^2)=\frac{2 m_T}{4\pi w^2}\left [ \frac{1}{2 j_{\chi}+1} \frac{1}{2 j_{T}+1}|\mathcal{M}_T(q^2)|^2 \right ],
\label{eq:dsigma_de}
\end{equation}
where $m_T$ is the mass of the nuclear target and $w$ is the incoming WIMP speed in the target reference frame. 

Notice that in the limit $q\rightarrow0$ the differential cross section scales as $q^{-4}$ and, as a consequence, the total cross section $\sigma_T = \int dE_R \frac{d\sigma_T}{d E_R} = 2 m_T \int d(q^2) \frac{d\sigma_T}{d E_R}$ diverges.

\subsection{Direct detection}
\label{sec:dd}
In a direct detection experiment, the number of expected nuclear recoil events within visible energy, $E^{\prime}_1 \le E^{\prime} \le E^{\prime}_2$, is given by:
\begin{eqnarray}
R_{\rm DD} &=& M_{\rm exp} {\tau_{\rm exp}} \hspace{0.5mm} \left(\frac{\rho_\odot}{m_{\chi}}\right) 
\sum_{T} N_{T}  \int du \hspace{0.5mm} f(u) \hspace{0.5mm} u 
\hspace{0.5mm} 
\int^{E_R^{\rm max}}_{0} dE_R \hspace{0.6mm} \zeta_T(E_R,E_1^{\prime},E_2^{\prime}) \hspace{0.5mm} \frac{d\sigma_T}{dE_R},
\label{eq:DD_event}
\end{eqnarray}

\noindent where $E_R^{\rm max} = 2 \mu^2_{\chi T} u^2 / m_T$ and $\zeta_T$ indicates the response of detector that depends on the visible energy range, the energy resolution and the efficiency:

\begin{equation}
    \zeta_T=\int_{E_1^{\prime}}^{E_2^{\prime}} dE^{\prime}{\cal
   G}_T\left [E^{\prime},Q(E_R)E_R)\right]\epsilon(E^{\prime}) \rightarrow \epsilon(E_R)\Theta(E_{R,2}-E_R)\Theta(E_R-E_{R,1}).
    \label{eq:zt}
    \end{equation}
 
 In Eq.~(\ref{eq:DD_event}) $M_{\rm exp}$, $\tau_{\rm exp}$ and $N_T$ are the fiducial mass of the detector, live--time of data taking and the number of targets per unit mass in the detector, respectively. The local density of DM is given by $\rho_\odot$ for which we use the standard value $\rho_\odot = 0.3$ $\rm GeV/cm^{3}$~\cite{DD_LEWIN1996}. The function $f(u)$ represents the normalised speed distribution of the incoming WIMPs 
in the reference frame of the solar reference frame, as noticed before. The differential scattering cross section $\frac{d\sigma_T}{dE_R}$ is given by Eq.~\eqref{eq:dsigma_de}. The maximum recoil energy of a scattering event is $E_{\rm max} = 2 \mu^2_{\chi T} u^2 / m_T$, with $\mu_{\chi T}$ the WIMP-nuclear reduced mass. 

Two of the experiments that we will consider in Section~\ref{sec:analysis}  (PICO--60 (C$_3$F$_8$)~\cite{pico60_2019} and PICO--60 (CF$_3$I)~\cite{pico60_2015}) provide their results directly in terms of the true recoil energy $E_R$ and encode the energy range in the acceptance $\epsilon$ (see Appendix~\ref{app:experiments}). In this case $Q(E_R)$ =1, ${\cal G}_T=\delta ( E^{\prime}-E_R)$ and $E^{\prime}_{1,2}$ = $E_{R,1,2}$, and the response of the detector simplifies to the acceptance times a window function selecting the experimental energy bin, as shown in the last step of Eq.~(\ref{eq:zt}).  

Note that, due to the presence of the energy threshold, a DD experiment 
is sensitive only to the speed range $u \geq u^{\rm DD-min}_T = \sqrt{m_T E^{\rm th}_T / 2 \mu^2_{\chi T}}$. This corresponds to the momentum 
threshold ($q^{\rm th}_T = \sqrt{2 m_T E^{\rm th}_T}$), ensuring that the energy integration in  Eq.~\eqref{eq:DD_event} for a given $u$ does not diverge when the interaction is driven by a massless mediator (Eq.~\eqref{eq:H_long}).

\subsection{Capture in the Sun}
\label{sec:capture}

WIMPs can be gravitationally captured by the Sun, where they annihilate to produce a neutrino flux that can be observed by neutrino telescopes (NTs). 
In the following we will assume WIMP capture to be driven by elastic scattering events off nuclear targets, a process that quickly brings WIMPs in thermal equilibrium with the solar plasma~\cite{Blennow_2018}. In case of a massless mediator WIMP self interactions may provide an additional contribution both to capture and thermalization~\cite{capture_long_range_Chen_2015,capture_long_range_Gaidau_2021}. The size of such contribution is model--dependent, but in general is expected to enhance the capture rate, therefore making our limits conservative.
For a thermal WIMP the reference value for the average of the WIMP annihilation cross section to SM particles time velocity $\langle \sigma v\rangle$ which corresponds to the observed DM density is of the order of $\langle \sigma v\rangle_{\rm thermal}\simeq$ 2.2$\times$10$^{-26}$ cm$^3$ s$^{-1}$~\cite{Thermal_relic_Steigman:2012nb}, and is the same driving today the annihilation of WIMP particles in the halo of our galaxy if DM annihilation proceeds in $s$--wave, i.e. if 
$\langle \sigma v \rangle$ is not suppressed by the non--relativistic WIMP velocity $v$. With these assumptions  the present bounds on the annihilation rate $\Gamma_\odot$ from NTs imply that the equilibration time $\tau_\odot$ is much smaller than the age of the Sun $t_\odot$~\cite{griest_seckel_1986} and $\Gamma_\odot$ = $C_\odot/2$ (i.e., equilibrium between capture and annihilation is achieved, a quantitative discussion on this is provided in Appendix (A.4) of \cite{halo_independent_sogang_2023}). Assuming equilibrium the corresponding neutrino flux is completely determined by the capture rate which  (in the optically thin limit) is driven by the differential cross section of Eq.~\eqref{eq:dsigma_de} 
and can be written as (see~\cite{Gould:1987}):
\begin{eqnarray}
C_\odot &=& \left(\frac{\rho_\odot}{m_{\chi}}\right) \hspace{0.5mm} \int du \hspace{0.5mm} f(u) 
\hspace{0.5mm} \frac{1}{u} \int^{R_\odot}_0 dr \hspace{0.5mm} 4 \pi r^2 \hspace{0.5mm} w^2
\nonumber\\ &&
\times \hspace{0.5mm} \sum_{T} \eta_{T}(r) \hspace{0.7mm} 
\Theta(E^{\rm C}_{\rm max} - E^{\rm C}_{\rm min}) 
\int^{E^{\rm C}_{\rm max}}_{E^{\rm C}_{\rm min}} dE \hspace{0.5mm} \frac{d\sigma_T}{dE} \, .
\label{eq:cap_rate}
\end{eqnarray}
Here  $w=\sqrt{u^2 + v^2_{\rm esc}(r)}$ is the 
incoming WIMP speed at a radial distance $r$ from the solar center, and 
$v_{\rm esc}(r)$ is the local escape speed at $r$ varying in the range 620$\sim$1400 km/s 
from the surface to the center of the Sun.
For the number density $\eta_{T}(r)$ of different target nuclei in the Sun we adopt the Standard Solar Model AGSS09ph~\cite{solar_model_Serenelli2009}.  

The maximum energy $E^{\rm C}_{\rm max}$ deposited by a WIMP in a scattering process in the Sun 
and the minimum energy $E^{\rm C}_{\rm min}$ that the WIMP needs to lose in order to be captured are, respectively:
\begin{eqnarray}
E^{\rm C}_{\rm max} &=& 2 \mu^2_{\chi T} w^2 / m_T = 2 \mu^2_{\chi T} (u^2 + v^2_{\rm esc}(r)) / m_T \, ,
\label{eq:E_max}\\
E^{\rm C}_{\rm min} &=& \frac{1}{2} m_\chi u^2\, .
\label{eq:E_min}
\end{eqnarray}

\noindent In the equation above, $E^{\rm C}_{\rm min}$ represents the minimal energy transfer that brings a WIMP (with speed $u$ at infinity and speed $\sqrt{u^2 + v^2_{\rm esc}(r)}$ at the distance $r$ from the Sun's center) below the escape velocity $v_{esc}(r)$. For $u\rightarrow 0$ such minimal energy vanishes, implying a divergence in the energy integral of Eq~(\ref{eq:cap_rate}) when the differential cross section is driven by the effective Hamiltonian~(\ref{eq:H_long}).
Physically, this is equivalent to consider a WIMP as captured whenever the scattering process locks it into a bound orbit irrespective of the size of its aphelion, with the assumption that during the lifetime of the solar system it will continue to scatter off the nuclear targets in the Sun and will be eventually driven to its core, where its annihilation process can be probed by NTs. It is instead more realistic to assume that the gravitational disturbances far from the Sun put an upper cut $r_0$ on the size of the maximal aphelion of the initial bound orbit. 
While the contribution of such very large orbits is irrelevant for a contact interaction, for the cross section diverging at small $q$, the capture rate becomes sensitive to $r_0$ and diverges for $r_0\rightarrow\infty$. A WIMP with initial position $r$ needs a minimal speed $v_e(r)^2=v_{esc}(r)^2-v_{esc}(r_0)^2$ to reach the maximal distance $r_0$. In this case a WIMP on an initial unbound trajectory and speed $u$ at infinity will need to lose the minimal energy:  

\begin{equation}
    E^{\rm C}_{\rm min} \rightarrow \frac{1}{2} m_\chi (u^2 + v_{esc}(r_0)^2) = \frac{1}{2} m_\chi u^2 + E^{\rm C}_{\rm cut}\, ,
    \label{eq:jupiter_cut}
\end{equation}

\noindent to be locked into a bound orbit with aphelion $r_0$. In this way $E^{\rm C}_{\rm min}$ never vanishes and the integral of Eq.~(\ref{eq:cap_rate}) yields a finite result.  In the literature, $r_0$ is identified with the distance from the Sun to Jupiter~\cite{jupiter_cut_Kumar_2012,Liang:2013dsa, Guo:2013ypa}, corresponding to:

\begin{equation}
v_{\rm esc}(r_0) = v_{\rm cut} \simeq 18.5 \, {\rm km/s} \, .
\label{eq:v_cut}
\end{equation}

\noindent In this case capture of a WIMP in the Sun via the scattering process off a target $T$ is kinematically possible up to a maximum value of the asymptotic WIMP speed $u^{\rm C-max}_T$
that is determined from the condition $E^{\rm C}_{\rm max} = E^{\rm C}_{\rm min}$:
\begin{equation}
(u^{\rm C-max}_T)^2 = v^2_{\rm esc}(r)\frac{4 m_{\chi} m_T}{(m_{\chi} - m_T)^2} - 
v_{esc}(r_0)^2 \frac{(m_\chi+m_T)^2}{(m_{\chi} - m_T)^2} 
\, ,
\label{eq:u_C_max}
\end{equation}
which yields the usual expression $u^{\rm C-max}_T = v_{\rm esc}(r) \sqrt{\frac{4 m_{\chi} m_T}{(m_{\chi} - m_T)^2}}$ for $v_{esc}(r_0) \rightarrow 0$. The Heaviside step function in Eq.~\eqref{eq:cap_rate} 
ensures that $C_\odot = 0$ for $E^{\rm C}_{\rm max} \leq E^{\rm C}_{\rm min}$, i.e., for $u^2 \geq (u^{\rm C-max}_T)^2$.

\section{Halo--independent bounds with the single--stream method}
\label{sec:single_stream} 

In this Section we outline the single--stream method originally introduced in~\cite{Halo-independent_Ferrer2015} to determine halo--independent bounds that do not depend on the velocity distribution $f(u)$.
For the WIMP speed distribution in the Solar system one can assume Eq.~(\ref{eq:f_normalization}) in the form:
\begin{equation}
\int^{u_{\rm max}}_0 du \hspace{0.5mm} f (u) = 1 \, ,
\label{eq:f_u_norm}
\end{equation}
\noindent where $u_{\rm max}$ is the maximum possible value of the asymptotic WIMP speed. In particular, assuming that the DM particles in the halo are gravitationally bound to the Galaxy, 
$u_{\rm max} = u_{\rm esc} + v_\odot$, with $v_\odot$ =220 km/s the value of the galactic rotational velocity and $u_{\rm esc}$ the escape speed in the Galactic rest frame, both at the Sun's position. 
In our analysis we will take the reference value $u_{\rm esc} = 560$ $\rm km/s$~\cite{vesc_Smith2006, vesc_Piffl2013} implying $u_{\rm max}$ = $u_{\rm max}^{ref}$ = 780 $\rm km/s$. However, we will also discuss the effects of choosing larger values for $u_{\rm max}$, possibly including a contribution from an extragalactic WIMP flux~\cite{Herrera:2023fpq}. 

The number of WIMP--induced nuclear recoil events in a DD experiment (Eq.~(\ref{eq:DD_event})) or 
the WIMP capture rate in the Sun (Eq.~(\ref{eq:cap_rate})) can be written as:
\begin{equation}
R = \int^{u_{\rm max}}_0 du \hspace{0.5mm} f(u) \hspace{0.5mm} H(u) .
\label{eq:rate_H_u}
\end{equation}
For DD, 
\begin{eqnarray}
H(u) = H_{\rm DD}(u) &=& M_{\rm exp} {\tau_{\rm exp}} \hspace{0.5mm} \left(\frac{\rho_\odot}{m_{\chi}}\right) u \hspace{0.5mm} 
\sum_{T} N_{T} \hspace{0.6mm} 
\int_{E^{\rm th}_T}^{E_{\rm max}} dE \hspace{0.6mm} \hspace{0.5mm} 
\zeta_T \hspace{1mm} \frac{d\sigma_T}{dE} \, ,
\label{eq:DD_H_u}
\end{eqnarray}
and for capture, 
\begin{eqnarray}
H(u) = H_C(u) &=& \left(\frac{\rho_\odot}{m_{\chi}}\right) \hspace{0.5mm} 
\hspace{0.5mm} \frac{1}{u} \int^{R_\odot}_0 dr \hspace{0.5mm} 4 \pi r^2 \hspace{0.5mm} w^2
\nonumber\\ &&
\times \hspace{0.5mm} \sum_{T} \eta_{T}(r) \hspace{0.7mm} 
\Theta(E^{\rm C}_{\rm max} - E^{\rm C}_{\rm min}) 
\int^{E^{\rm C}_{\rm max}}_{E^{\rm C}_{\rm min}} dE \hspace{0.5mm} \frac{d\sigma_T}{dE} \, , 
\label{eq:NT_H_u}
\end{eqnarray}
\noindent with $\frac{d\sigma_T}{dE}$ driven by the long--range interaction \eqref{eq:H_long}.

Considering a WIMP--nucleon coupling $\alpha$, a given experimental bound $R_{\rm max}$ (either from a DD or a NT experiment) implies:
\begin{equation}
R = R(\alpha^2) = \int^{u_{\rm max}}_0 du f(u) H(\alpha^2, u) \leq R_{\rm max} \, ,
\label{eq:rate1}
\end{equation}
with $H$ being either $H_C$ or $H_{\rm DD}$. Since $H(\alpha^2, u) = \alpha^2 H(\alpha = 1, u)$, one can re--write (\ref{eq:rate1}) as:
\begin{equation}
R(\alpha^2) = \int^{u_{\rm max}}_0 du f(u) \frac{\alpha^2}{{\alpha^2}_{\rm max}(u)} 
H({\alpha^2}_{\rm max}(u), u) = 
\int^{u_{\rm max}}_0 du f(u) \frac{\alpha^2}{{\alpha^2}_{\rm max}(u)} R_{\rm max} \leq R_{\rm max} \, ,
\label{eq:rate2}
\end{equation}
where ${\alpha}_{\rm max}(u)$ is defined as:
\begin{equation}
H({\alpha^2}_{\rm max}(u), u) = {\alpha^2}_{\rm max}(u) H(\alpha=1, u) = R_{\rm max} \, ,
\label{eq:c_max_u}
\end{equation}
i.e., ${\alpha}_{\rm max}(u)$ is the upper limit on the coupling $\alpha$ if all the incoming WIMPs were concentrated in a single stream with speed $u$. Using Eq.~(\ref{eq:rate2}) one obtains the upper limit on the  coupling $\alpha$ for a general WIMP speed distribution $f(u)$ as:
\begin{equation}
\alpha^2 \leq \left[\int^{u_{\rm max}}_0 du \frac{f(u)}{{\alpha^2}_{\rm max}(u)}\right]^{-1} .
\label{eq:c2_upper_bound}
\end{equation}

According to Eq.~\eqref{eq:c2_upper_bound}, a finite bound on the coupling $\alpha$ 
independent of $f(u)$ (i.e., a finite halo--independent bound) 
is possible if the experimental sensitivity extends to the full WIMP speed range, so that ${\alpha}_{\rm max}(u)<\infty$ for any $u \in [0, u_{\rm max}]$.  Since DD searches are sensitive to $u \geq u^{\rm DD-min}$ and 
NTs to $u \leq u^{\rm C-max}$, only the combination of both is sensitive to the full range of WIMP speeds $u^{\rm C-max} \geq$ min$\left[u^{\rm DD-min}, u_{\rm max}\right]$. If this happens, one of the two following situations can occur: 

\begin{itemize}
\item \underline{Case I:}
The NT and the DD experiments are sensitive to two specific ranges of the WIMP speed, 
giving:
\begin{eqnarray}
{({\alpha^{\rm NT}})^2}_{\rm max}(u) &\leq& \tilde{\alpha}^2 
\hspace{18mm} {\rm for} \hspace{2mm} 0 \leq u \leq \tilde{u}
\nonumber\\
{({\alpha^{\rm DD}})^2}_{\rm max}(u) &\leq& \tilde{\alpha}^2 
\hspace{18mm} {\rm for} \hspace{2mm} \tilde{u} \leq u \leq u_{\rm max}
\label{eq:condition_I}
\end{eqnarray}
where ${{\alpha^{\rm NT}}}_{\rm max}(u)$ and ${{\alpha^{\rm DD}}}_{\rm max}(u)$ correspond to ${\alpha}_{\rm max}(u)$ for the NT and the DD experiments, respectively, 
and $\tilde{u}$ indicates the speed where 
${{\alpha^{\rm NT}}}_{\rm max}(u)$ and ${{\alpha^{\rm DD}}}_{\rm max}(u)$ 
intersect at a finite value $\tilde{\alpha}$, 
i.e., ${{\alpha^{\rm NT}}}_{\rm max}(\tilde{u}) = {{\alpha^{\rm DD}}}_{\rm max}(\tilde{u}) 
= \tilde{\alpha}$. In this case, using Eq.~(\ref{eq:c2_upper_bound}) it can be shown that 
(see \cite{Halo-independent_Ferrer2015} and \cite{halo_independent_sogang_2023} for details) 
the halo--independent (HI) bound on the coupling is:
\begin{equation}
\alpha^2 \leq 2 \hspace{0.5mm} \tilde{\alpha}^2 = \alpha^2_{\rm HI} .
\label{eq:limit_I}
\end{equation}

\item \underline{Case II:} 
In some cases it happens that:
\begin{equation}
{({\alpha^{\rm DD}})^2}_{\rm max}(u) > \tilde{\alpha}^2 \hspace{18mm} {\rm at} 
\hspace{2mm} u = u_{\rm max} .
\label{eq:condition_II}
\end{equation}
In this case, as shown in \cite{halo_independent_sogang_2023}, the halo--independent bound is:
\begin{equation}
\alpha^2 \leq {({\alpha^{\rm DD}})^2}_{\rm max}(u_{\rm max}) + \tilde{\alpha}^2 = \alpha^2_{\rm HI} .
\label{eq:limit_II}
\end{equation}
The condition~\eqref{eq:condition_II} occurs mainly because of the loss of the DD experimental sensitivity with increasing $u$, usually due to the suppression of the 
nuclear form--factor in the WIMP--nucleus scattering process  at large momentum transfer $q$. Note that this condition implies that the halo--independent bound 
(obtained from Eq.~\eqref{eq:limit_II}) becomes sensitive to the choice of $u_{\rm max}$ and 
increases with increasing $u_{\rm max}$. 
In order to see the quantitative effect of this in our analysis we 
will also consider a large value of $u_{\rm max} = 8000$ km/s 
(which is an order of magnitude larger than $u_{\rm max}^{ref} = 780$ km/s).

\end{itemize}

For a given WIMP--nucleon interaction coupling, the halo--independent upper bound 
is computed at each $m_{\chi}$ following 
either Eq.~(\ref{eq:limit_I}) or (\ref{eq:limit_II}) (when appropriate) combining one DD with one NT. 
When more than one DD and one NT are involved (as in our analysis) 
the procedure is repeated for each combination of DD and NT, 
and the most constraining halo--independent limit on the coupling is taken.

\section{Analysis}
\label{sec:analysis}

We begin this Section by reviewing the bounds on the SI and SD dark matter
couplings $\alpha_{SI}^{p,n}$ and $\alpha_{SD}^{p,n}$ when the SHM is assumed for the WIMP velocity distribution in our Galaxy. In such case we assume the following normalised distribution 
(see e.g., \cite{Gould:1987, jupiter_cut_Kumar_2012}): 
\begin{equation}
f_{\rm MB}(u) = 2 \sqrt{\frac{3}{2\pi}} \left(\frac{u}{v_\odot v_{\rm rms}}\right) 
{\rm exp}\left[-\frac{3(u^2+v^2_\odot)}{2v^2_{\rm rms}}\right] 
{\rm sinh}\left(\frac{3uv_\odot}{v^2_{\rm rms}}\right)\, ,
\label{eq:f_MB}
\end{equation}
with rotational speed of the solar system $v_\odot = 220$ $\rm km/s$ and 
velocity dispersion $v_{\rm rms} = \sqrt{3/2} \, v_\odot \simeq 270$ km/s~\cite{SHM_maxwell_Green2011}.

\begin{figure*}[ht!]
\centering
\includegraphics[width=7.49cm,height=6.3cm]{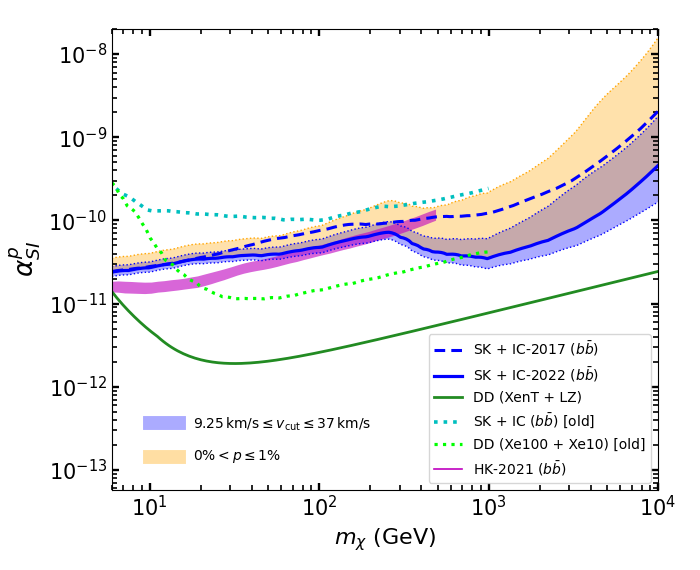}
\includegraphics[width=7.49cm,height=6.3cm]{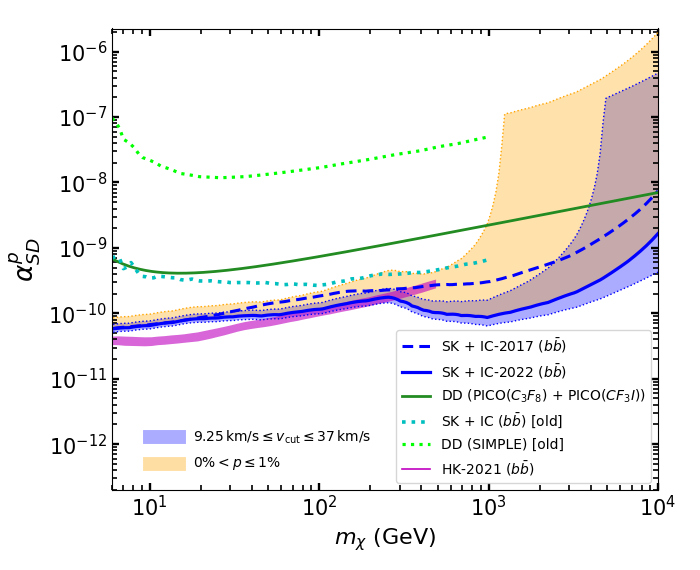}\\\includegraphics[width=7.49cm,height=6.3cm]{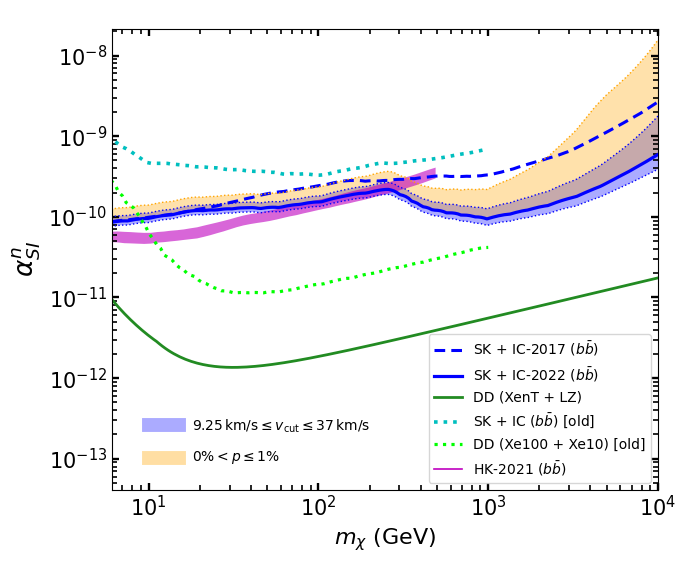} 
\includegraphics[width=7.49cm,height=6.3cm]{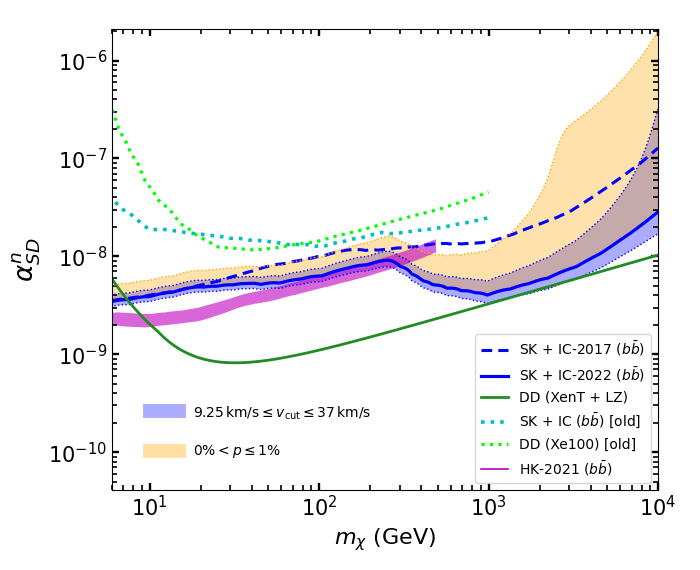}
\caption{An update of the bounds on the long--range SI and SD interaction couplings considering the Standard halo model (SHM). The old bounds on such interactions 
(extracted from \cite{Guo:2013ypa} considering the SI interaction and \cite{Liang:2013dsa} for the SD interaction for both NT and DD) are shown by dotted curves. 
The future projections from the upcoming Hyper--Kamiokande \cite{Bell:2021esh} 
are shown by the purple bands.}
\label{fig:coupling_mx_SHM}
\end{figure*}

Our results are shown in Fig.~\ref{fig:coupling_mx_SHM}, where we report for reference the results from Refs.~\cite{Guo:2013ypa} and~\cite{Liang:2013dsa}, which are the most recent analyses on the subject previous to the present paper. Specifically, the bounds from capture in the Sun assume $b\bar{b}$ as the dominant WIMP annihilation channel and are shown with the blue dotted line and the DD bounds  with the green dotted lines, respectively. 
At the time of the analyses of Refs.~\cite{Guo:2013ypa} and~\cite{Liang:2013dsa} the most stringent DD bounds on all the couplings $\alpha_{SI}^{p,n}$ and $\alpha_{SD}^{p,n}$ of Eq.~(\ref{eq:H_long}) were from Xenon10~\cite{xenon10_2011} and Xenon100~\cite{xenon100_2012}, while for capture in the Sun they were from Super--Kamiokande~\cite{superk_2011} and IceCube~\cite{icecube_2011,icecube_2012}. In the same Figure 
our evaluation of the present combined NT limits from the latest Super--Kamiokande~\cite{SuperK_2015} and IceCube~\cite{IceCube:2016} data is shown by the dashed blue lines, with the solid blue lines representing the improvement using a more recent set of IceCube data presented at the ICRC2021 Conference~\cite{IceCube:2022wxw}. Moreover in the same plots we show separately with the green solid lines our updated estimation of the present DD limit, which are given by a combination of LZ~\cite{LZ_2022} and XenonT~\cite{xenon_nT} for $\alpha_{SI}^{p,n}$ and $\alpha_{SD}^{n}$, and by a combination of PICO--60 (C$_3$F$_8$)~\cite{pico60_2015} PICO--60 (CF$_8$I)~\cite{pico60_2019} for $\alpha_{SD}^{p}$.  

From Fig.~\ref{fig:coupling_mx_SHM} one can notice that, albeit the present combined bounds from DD and capture in the Sun on $\alpha_{SI}^{p,n}$ and $\alpha_{SD}^{p,n}$ have improved between about two and three orders of magnitude since the analysis of Refs.~\cite{Guo:2013ypa,Liang:2013dsa}, the situation has remained qualitatively similar: in the case of the SI couplings $\alpha_{SI}^{p,n}$ and of the SD coupling $\alpha_{SD}^{n}$ the bounds are driven by DD, while in the case of $\alpha_{SD}^{p}$ the most stringent constraint comes from capture in the Sun. One also observes that a long--range interaction systematically enhances the signal in NTs compared to that in DD due to the fact that, for a massless mediator, events with a small momentum transfer enhance the WIMP--nucleus cross section in the capture rate in the Sun while are not observable in DD due to the experimental energy threshold. As a consequence, while for a contact interaction the sensitivity of DD is always better than that for capture in the Sun in the case of a $b\bar{b}$ final state (between two and three orders of magnitude for $c_{SI}^{p,n}$ and $c_{SD}^{n}$, and about one order of magnitude for $c_{SD}^{p}$, see \cite{halo_independent_sogang_2023}), for a long--range interaction, thanks to the large amount of $^1H$ in the Sun, the bound on $\alpha_{SD}^{p}$ is driven by capture, while for the other couplings DD is more sensitive but the difference with capture reduces in a sizeable way. It is also worth noticing that at the time of~\cite{Guo:2013ypa,Liang:2013dsa} the bound on WIMP--proton couplings from SIMPLE~\cite{simple_2012}, that used C$_2$CIF$_5$ and was the most sensitive experiments using proton--odd targets, was slightly worse than the corresponding one from Xenon100, that uses a neutron--odd target. However the Xenon100 constraint  required to use sub--dominant WIMP--proton form factors that show significant variability among different evaluations~\cite{SD_form_factors_xe_1, SD_form_factors_xe_2,nreft_haxton2}. On the other hand the present bound on $\alpha_{SD}^{p}$ from PICO--60, and experiments using proton--odd targets (C$_3$F$_8$ and CF$_3$I) should be considered more robust since it makes use of better known SD form factors. 

In the case of a massless mediator the response functions $H_C$, $H_{DD}$ are enhanced for a low momentum transfer. Due to the experimental threshold this effect is mild or negligible for DD experiments, but can be important for the expected rate of capture in the Sun, that for $u\rightarrow 0$ diverges when the integration over the nuclear recoil energy in Eq.~(\ref{eq:cap_rate}) extends down to zero, unless the Jupiter cut--off discussed in Section~\ref{sec:capture} is implemented (see Eqs.~(\ref{eq:E_min}) and (\ref{eq:jupiter_cut})). 

One can notice that the rate of Eq.~(\ref{eq:rate_H_u}) can be seen as the average $\langle$H$\rangle$ of the response function $H$ over the velocity distribution $f$. Indeed, for a constant $H$ the rate would not depend on $f$ altogether. At low WIMP masses, although $H_C$ is never constant ($H_C(u)\propto 1/u$ or $H_C\propto 1/u^3$ depending on whether $E_{cut}^C>1/2 m_\chi u^2$ or not in Eq.~(\ref{eq:jupiter_cut})) the response function $H_C$ flattens out, because $u_T^{C-max}>u_{\rm max}$ so that $H_C(u)\ne$ 0 in the full velocity range, and this reduces the sensitivity of the capture signal on the details of the velocity distribution. On the other hand, at larger masses $u_T^{C-max}<u_{\rm max}$ and $H_C(u)$ has a steeper $u$ dependence because it vanishes for $u>u_T^{C-max}$. This leads to a larger sensitivity of the capture rate on the WIMP velocity distribution at heavier WIMP masses, and specifically on the low--speed regime of $f(u)$~\cite{Choi:2013eda}. For the SHM this is explicitly shown in Fig.~\ref{fig:coupling_mx_SHM}, where in each plot the blue shaded band represents the change in the bound from capture in the Sun when the $v_{esc}(r_0)$ value of Eq.~(\ref{eq:v_cut}) is multiplied/divided by a factor of two (corresponding to assuming a maximal semi-major axis  of captured WIMPs a factor of four 
smaller/larger than the Sun-Jupiter distance). In the same plots the yellow band represents instead how the same bound is modified when the tail $u<u_p$ of the standard Maxwellian distribution $f_{MB}$ is removed from the velocity integration, with:

\begin{eqnarray}
    &&p\equiv \int_0^{u_p} f_{MB}(u)\,du = 0.01\nonumber\\
    &&\rightarrow u_p\simeq71\,\mbox{km/s}. 
\end{eqnarray}

\noindent Indeed one can observe that the effect is mild (a factor of a few) for $m_\chi\lesssim$ 100 GeV, but  becomes much more sizeable at larger masses. 

\begin{figure*}[ht!]
\centering
\includegraphics[width=7.49cm,height=6.3cm]{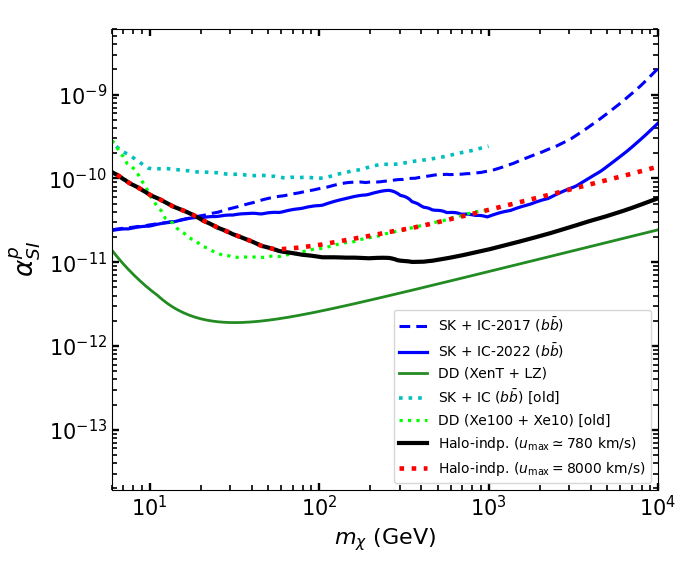}
\includegraphics[width=7.49cm,height=6.3cm]{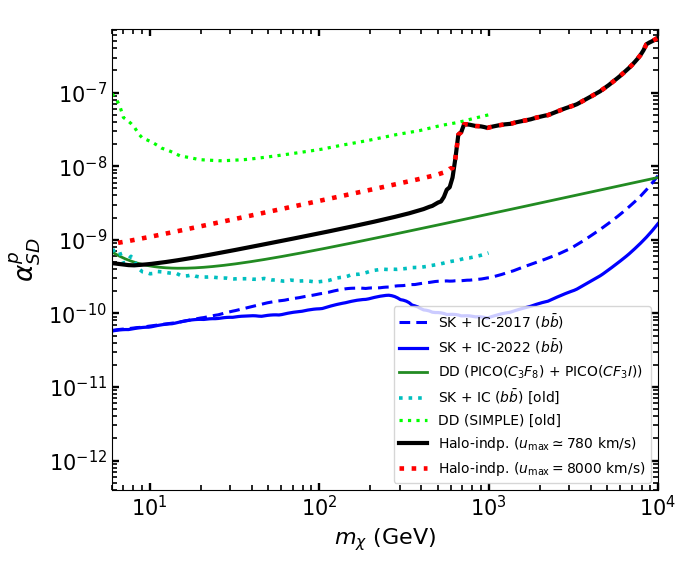}\\\includegraphics[width=7.49cm,height=6.3cm]{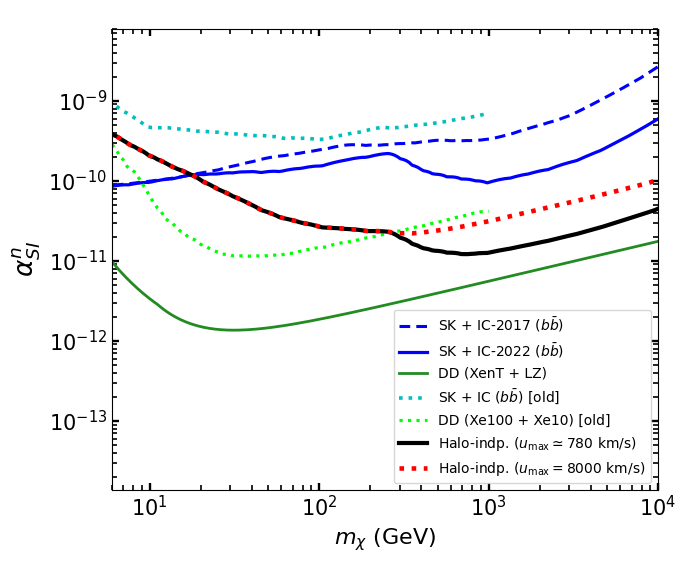} 
\includegraphics[width=7.49cm,height=6.3cm]{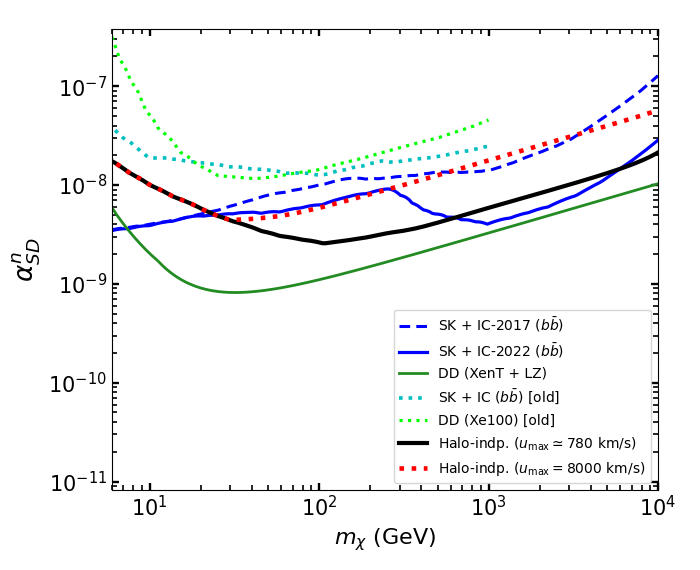}
\caption{Halo--independent bounds are shown by the black solid lines 
(considering $u_{\rm max}$ = $u_{\rm max}^{ref}$ = 780 $\rm km/s$). 
Similar bounds taking $u_{\rm max}$ = 8000 $\rm km/s$ are shown by the red dotted lines. 
For comparison, the SHM bounds of Fig.~\ref{fig:coupling_mx_SHM} are also shown.}
\label{fig:coupling_mx_HI}
\end{figure*}

We now extend our analysis beyond the SHM to the halo--independent constraints obtained adopting the single--stream method described in Section~\ref{sec:single_stream}, which is valid for any velocity distribution with the condition~(\ref{eq:f_u_norm}). Also in this case we fix the WIMP annihilation channel to $b\bar{b}$. In each plot of Fig.~\ref{fig:coupling_mx_HI}, the solid black line shows our halo--independent result for each of the couplings $\alpha_{SI,SD}^{p,n}$ as a function of $m_\chi$ and using for the maximal velocity $u_{\rm max}$ = $u_{\rm max}^{\rm ref}$ = 780 km/s, corresponding to the standard value $u_{\rm esc}$ = 560 km/s for the  escape velocity in the Galactic rest frame~\cite{vesc_Piffl2013,vesc_Smith2006}. 
As pointed out in Section~\ref{sec:single_stream}, in some cases the halo--independent bound can become sensitive to the choice of $u_{\rm max}$. Such situations are described by case II (see Eq.~\eqref{eq:condition_II}) when at
large WIMP speeds the DD response function $H_{\rm DD}(u)$ is suppressed, or equivalently, 
the DD bound on the coupling, ${\alpha^{\rm DD}}_{\rm max}(u)$, is relaxed. 
This happens due to the suppression of the nuclear form factor at large momentum transfer (corresponding to large WIMP speeds) and/or due to the finite experimental energy bin of some detectors (e.g., xenon based detectors). 
As a result of this the halo--independent bound obtained from Eq.~\eqref{eq:limit_II} in case II gets weaker when a large value of $u_{\rm max}$ is considered. For the same reason described above, increasing $u_{\rm max}$ can lead to a transition from  
a situation described by case I (see Eq.~\eqref{eq:condition_I}) to one described by case II (Eq.~\eqref{eq:condition_II}) enhancing the sensitivity of the bound to $u_{\rm max}$. 
To check the effects of choosing a larger $u_{\rm max}$ on the halo--independent bounds 
in each plot of Fig.~\ref{fig:coupling_mx_HI} the red--dotted line shows the bound when $u_{\rm max}$ is increased to 8000 km/s. Such value is unrealistically large, and we use it just to show that the dependence on $u_{\rm max}$ is rather mild (the halo--independent bounds get weaker only by a factor $\lesssim 2-3$).

\begin{figure*}[ht!]
\centering
\includegraphics[width=7.49cm,height=6.2cm]{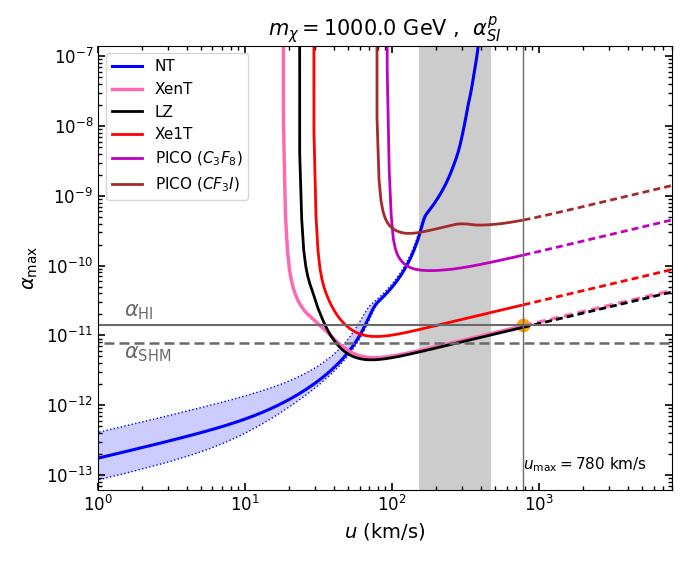}
\includegraphics[width=7.49cm,height=6.2cm]{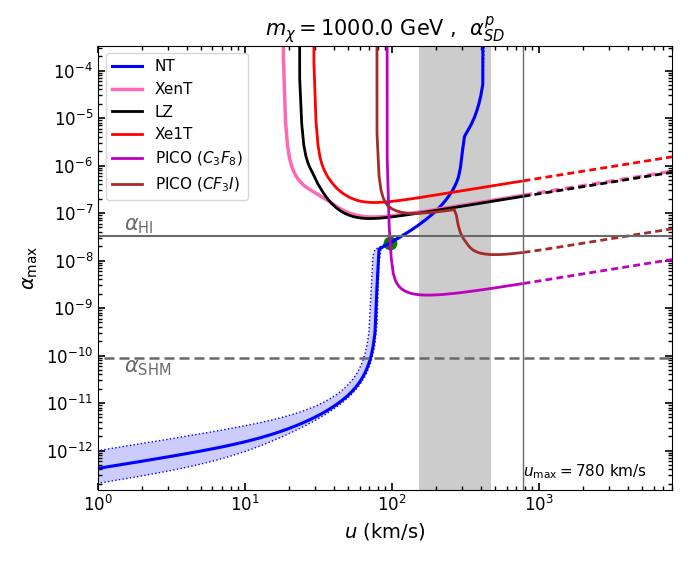}
\includegraphics[width=7.49cm,height=6.2cm]{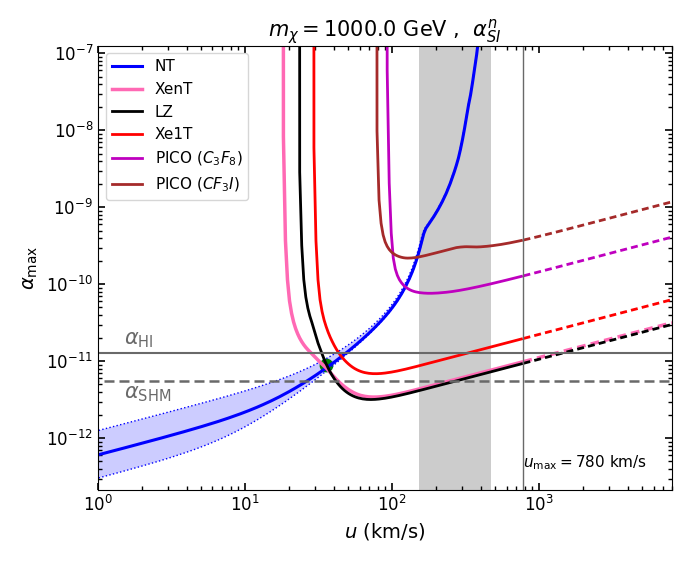}
\includegraphics[width=7.49cm,height=6.2cm]{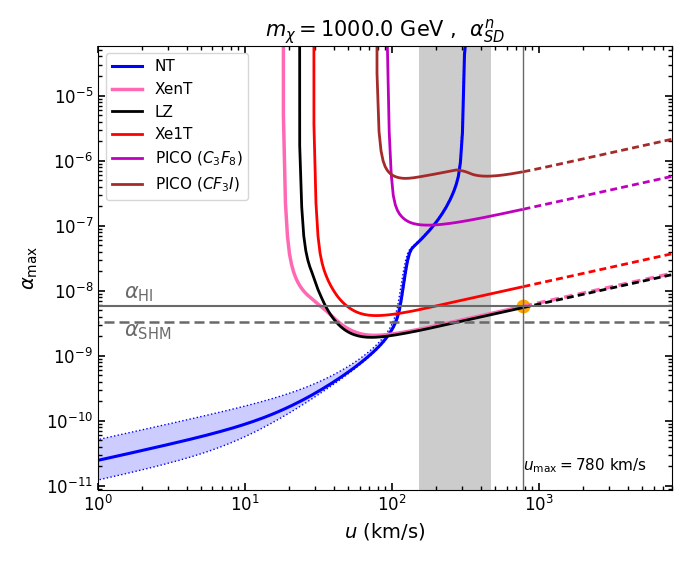}
\caption{${\alpha}_{\rm max}$ (defined in Eq.~(\eqref{eq:c_max_u})) as a function of the WIMP stream speed $u$ shown for different long--range interaction couplings considering different NT and DD experiments and for $m_\chi$ = 1 TeV. The halo--independent (HI) bound 
(obtained taking $u_{\rm max}$ = $u_{\rm max}^{ref}$ = 780 $\rm km/s$) 
on each coupling is indicated by the horizontal solid line, while the 
corresponding SHM bound is shown by the horizontal dashed line. 
The blue shaded region in each case corresponds to the change in $v_{\rm cut}$ 
by multiplying/dividing $v_{esc}(r_0)$ by a factor of two. 
The gray shaded region in each case indicates the range of $u$ for 
which $\int f_{\rm MB}(u) du \simeq 0.8$.}
\label{fig:copulingMax_u}
\end{figure*}

An explicit example of how the bounds of Fig.~\ref{fig:coupling_mx_HI} are obtained is shown in Fig.~\ref{fig:copulingMax_u} for $m_\chi$ = 1 TeV, where the quantity ${\alpha}_{\rm max}$ defined in Eq.~(\ref{eq:c_max_u}) is plotted as a function of $u$. In each plot the green or orange circle represents the values $(\tilde{u}, \tilde{\alpha})$ or $(u_{\rm max},{\alpha^{\rm DD}}_{\rm max}(u_{\rm max}))$ that determine the single--stream bound according to the procedure outlined in Section~\ref{sec:single_stream} (case I or II, respectively), the gray shaded area indicates the range of $u$ preferred by the SHM (specifically, the $u$ interval where $\int f_{\rm MB}\,du=0.8$), while the vertical solid line represents the $u_{\rm max}$ reference value $u_{\rm max}^{\rm ref}$ = 780 km/s. An important feature of these plots is that the effect of the Jupiter cut on the determination of $\alpha_{\rm max}(u)$ is confined to $u\rightarrow 0$, where $\lim_{u\rightarrow 0}\alpha_{\rm max, cut}>0$ and $\lim_{u\rightarrow 0}\alpha_{\rm max}$ = 0 without the cut. However, the halo--independent bound is determined by the {\it largest} value of $\alpha_{\rm max}(u)$ across the full range of $u$ and is insensitive to $u$ intervals where $\alpha_{\rm max}(u)$  is small. As a consequence, the halo--independent bound does not depend on the Jupiter cut, as it should be. Moreover, for $m_\chi$ = 1 TeV one can notice that the couplings $\alpha_{SD}^p$ and $\alpha_{SI}^n$ correspond to case I (as defined in Section~\ref{sec:single_stream}), for which the halo--independent bound is given by Eq.~(\ref{eq:limit_I}) with small or no dependence on $u_{\rm max}$. On the other hand the other two plots in Fig.~\ref{fig:copulingMax_u}  show that $\alpha_{SI}^p$ and $\alpha_{SD}^n$  correspond to case II, for which $\tilde{u}$ = $u_{\rm max}$ and the single--stream bound is given by Eq.~(\ref{eq:limit_II}). In this case some dependence on the specific value of $u_{\rm max}$ is expected, and this is confirmed by the red--dotted lines at large mass in the corresponding plots of Fig.~\ref{fig:coupling_mx_HI}, conservatively calculated with the very large value $u_{\rm max}$ = 8000 km/s to  bracket  the ensuing variation of the bound. The two cases $\alpha_{SI}^{p,n}$ are similar, in that the cross section scales as the atomic mass number and DD is driven by xenon detectors, while capture in the Sun is driven by WIMP scattering events off several nuclear targets 
(with the largest from either $^1H$ or $^4He$, followed by $^3He$, $^{12}C$, $^{14}N$, $^{16}O$, $^{20}Ne$, $^{24}Mg$, $^{28}Si$ and $^{56}Fe$).
In this case the maximal velocity for capture $u^{\rm C-max}_T$ for several targets extends well beyond the velocity thresholds $u \geq u^{\rm DD-min}_T$ from direct searches, driven by xenon. The same happens for $\alpha_{SD}^{n}$, for which capture in the Sun is driven by WIMP scattering events off $^3He$ and, again, DD is driven by xenon detectors. In all these cases DD and capture keep a similar degree of complementarity and the halo--independent bound variation is moderate across the full range of $m_\chi$. On the other hand, the situation for $\alpha_{SD}^{p}$ is different, because in this case the capture rate in the Sun and DD are driven by WIMP scattering events off hydrogen and off fluorine targets, respectively. In particular, for $m_\chi \gtrsim$ 1 TeV 
$u^{\rm C-max}_{^H}$ for hydrogen drops below $u^{\rm DD-min}_F$
for fluorine and the complementarity between these two targets is lost. The residual contribution to capture beyond 
$u^{\rm C-max}_{^H}$ 
is driven by $^{14}N$, which is more than 6 orders of magnitude smaller. This explains the abrupt loss of sensitivity observed for $m_\chi\gtrsim$ 1 TeV in the halo--independent $\alpha_{SD}^{p}$ exclusion plot of  Fig.~\ref{fig:coupling_mx_HI}.

\begin{figure*}[ht!]
\centering
\includegraphics[width=7.49cm,height=6.3cm]{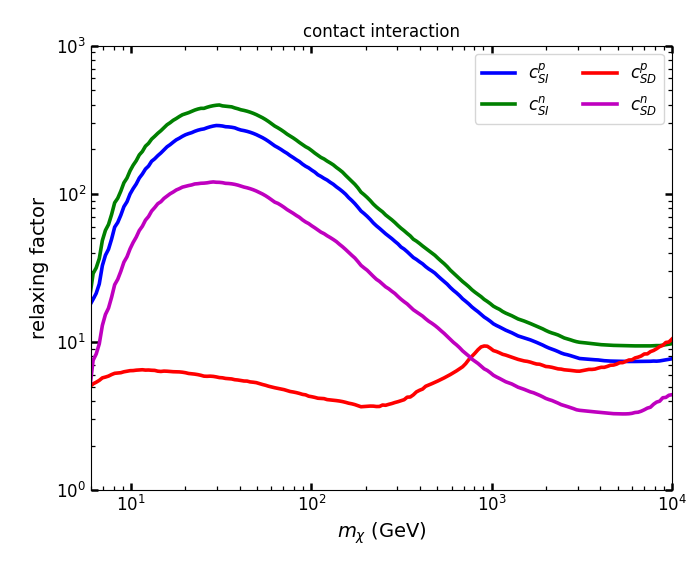}
\includegraphics[width=7.49cm,height=6.3cm]{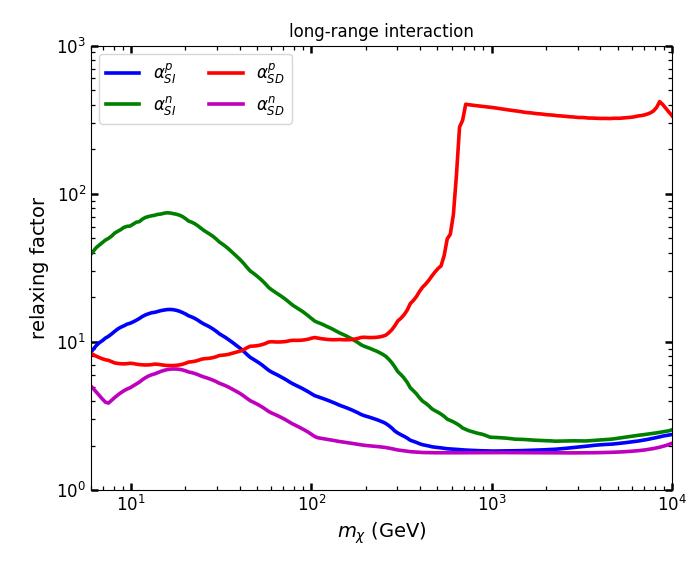}
\caption{Relaxation factor defined as the ratio between the halo--independent bound and the SHM one. Left--hand plot: contact interaction; right--hand plot: long--range interaction (massless propagator).}
\label{fig:relaxing_factor}
\end{figure*}

We conclude our discussion by plotting in Fig.~\ref{fig:relaxing_factor}  as a function of $m_\chi$ a relaxation factor defined as the ratio between 
the following two quantities: the conservative halo--independent exclusion plot obtained using the procedure outlined in Section~\ref{sec:single_stream} and the strongest constraint among capture in the Sun and the DD experiments included in our analysis obtained using for $f(u)$ the standard Maxwellian distribution $f_{MB}(u)$ of Eq.~(\ref{eq:f_MB})~\footnote{As already pointed out, in our analysis we assume $b\bar{b}$ as the dominant WIMP annihilation channel. For completeness, in Appendix~\ref{app:NT} we provide a plot of the relaxation factor of Fig.~\ref{fig:relaxing_factor} for the case of the $\tau\bar{\tau}$ annihilation channel.}. For the sake of comparison in Fig.~\ref{fig:relaxing_factor}  the left--hand plot shows the relaxation factors for $c_{SI,SD}^{p,n}$ in the case of a contact interaction (taken from Ref.~\cite{halo_independent_sogang_2023}), while the right--hand plot shows the relaxation factors for the couplings $\alpha_{SI,SD}^{p,n}$ in the case of a long--range interaction (i.e. for a massless mediator). Such relaxation factors show the maximal weakening of each bound when a halo model different from the SHM is assumed, and can be considered as indicators of the impact on the bounds of our ignorance about the velocity distribution $f(u)$. However, one should notice that, as shown in Fig.~\ref{fig:coupling_mx_SHM},  the SHM bound on the $\alpha_{\rm SD}^p$ coupling is driven by capture, which at large WIMP masses is very sensitive to the specific behaviour of $f_{\rm MB}$ for $u\rightarrow$ 0. 
This implies that for such bound the exact determination of the relaxation factor becomes not reliable, i.e., for $m_\chi\gtrsim$ 1 TeV. On the other hand all the other SHM bounds (determined by DD) and that on $\alpha_{\rm SD}^p$ for $m_\chi\lesssim$ 1 TeV are not sensitive to the low velocity tail of $f_{\rm MB}$ and the corresponding relaxation factors are robust. Interestingly, in this case the relaxation factor shows values which are smaller or of the same order of those for a contact interaction. This can be ascribed to the fact that the relaxation factor is explicitly given by~\cite{halo_independent_sogang_2023}: 
\begin{equation}
r_f^2 = \frac{\alpha^2_{\rm HI}}{(\alpha^{\rm exp}_{\rm SHM})^2} = \alpha^2_{\rm HI}  \int^{u_{\rm max}}_0 du 
\frac{f_{M}(u)}{({\alpha^{\rm exp}})^2_{\rm max}(u)} = 
\alpha^2_{\rm HI} \hspace{1mm} {\left \langle \frac{1}{({\alpha^{\rm exp}})^2_{\rm max}} \right \rangle}
\simeq \alpha^2_{\rm HI}  \hspace{1mm} {\left \langle \frac{1}{({\alpha^{\rm exp}})^2_{\rm max}} \right \rangle}_{\rm bulk}\, ,
\label{eq:rf}
\end{equation}
where ``exp" indicates the NT or DD experiment that 
provides the strongest upper--limit on the coupling at a given $m_{\chi}$ in the case of the standard Maxwellian speed distribution $f_{MB}$. 
The brackets $\langle ...\rangle$ indicate an average weighted by the Maxwellian $f_{MB}$ while  ${\langle ...\rangle}_{\rm bulk}$ indicates the dominant contribution to the average from the bulk of the WIMP speeds, defined as:
\begin{equation}
\int_{\rm bulk} du f_{M}(u) \simeq 1-\epsilon,
\label{eq:bulk}
\end{equation}

\noindent with $\epsilon$ some small number. With the exception of $\alpha_{\rm SD}^p$ at large WIMP masses ($m_\chi\gtrsim$ 1 TeV), in Eq.~(\ref{eq:rf}) the velocity interval which determines both $\alpha^2_{\rm HI}$ and the average ${\left \langle \frac{1}{({\alpha^{\rm exp}})^2_{\rm max}} \right \rangle}_{\rm bulk}$ is far from the $u\rightarrow$ 0 regime responsible for the sensitivity of the expected capture rate for a massless propagator (such regime corresponds qualitatively to where the thickness of the blue shaded regions in Fig.~\ref{fig:copulingMax_u}, representing  the change in  $\alpha_{\rm max}$ for capture when $v_{esc}(r_0)$ is multiplied/divided by a factor of 2, becomes sizeable). As a consequence, the discussion of the relaxation factor's behaviour for a long--range interaction is similar to the case of a contact interaction~\cite{halo_independent_sogang_2023}. 

At small $m_\chi$, both $u^{\rm C-max}$ and $u^{\rm DD}_{th}$ are shifted to large values, so that $(\alpha^{\rm NT})^2_{\rm max}(u)\lesssim(\alpha^{\rm DD})^2_{\rm max}(u)$  with $(\alpha^{\rm NT})^2_{\rm max}(u)$ rather flat up to $u_{\rm max}$, while $\tilde{u}$ is beyond the Maxwellian bulk region or close to its upper edge. In this case $(\alpha^{\rm NT})^2_{\rm max}(u)$ remains flat in a speed range that includes both the bulk of the Maxwellian and $\tilde{u}$ and as a consequence in Eq.~(\ref{eq:rf}) $\alpha^2_{\rm HI}$ 
and $(\langle 1/(\alpha^{\rm NT})^2_{\rm max} \rangle)^{-1}$
do not differ much, so that the relaxation factor is not large.

In the opposite regime of large $m_\chi$, both $u^{\rm DD}_{th}$ and $u^{\rm C-max}$ are shifted to small values and in the Maxwellian case the bound is driven by DD. In this case $\tilde{u}$ is also driven to small values and below the bulk of the Maxwellian, where $(\alpha^{\rm NT})^2_{\rm max}(u)$ intersects $(\alpha^{\rm DD})^2_{\rm max}(u)$ 
just before the latter starts rising due to the $u^{\rm DD}_{th}$ threshold. In this case the range of speeds that includes the bulk of the Maxwellian and either $\tilde{u}$ or $u_{\rm max}$ (case I or case II) corresponds to a regime where $(\alpha^{\rm DD})^2_{\rm max}(u)$ is rather flat. So also in this case in Eq.~(\ref{eq:rf}) the difference between $\alpha_{\rm HI}^2$ and $(\langle 1/(\alpha^{\rm DD})^2_{\rm max} \rangle)^{-1}$ is not large and the relaxation factor is moderate.

Finally, between the two asymptotic regimes of moderate relaxation factor at both small and large $m_\chi$ discussed above, for all the couplings $\alpha_{\rm SI,SD}^{p,n}$ 
the largest values of the relaxation factors are reached for $m_\chi\simeq$ 20 GeV
\footnote{At least at face value: as already pointed out, due to the uncertainty in the Maxwellian capture rate the relaxation factor for $\alpha_{\rm SD}^{p}$ is not well defined for large $m_\chi$.}.
In this mass range when the Maxwellian bound is driven by DD the two curves for $(\alpha^{\rm NT})^2_{\rm max}(u)$ 
and $(\alpha^{\rm DD})^2_{\rm max}(u)$ intersect where the latter has a steep dependence on $u$ because $u$ is close 
to $u^{\rm DD}_{th}$, with $\tilde{u}$ close to the lower edge of the bulk region. Due to these reasons in the range of speeds that includes $\tilde{u}$ 
and the bulk of the Maxwellian $(\alpha^{\rm DD})^2_{\rm max}(u)$ changes significantly, so that
$(\alpha^{\rm DD})^2_{\rm max}(u) \ll \tilde{\alpha}^2$ for WIMP speeds in the bulk of the Maxwellian, and the relaxation factor is large. Notice that in the case of $\alpha_{\rm SD}^p$ the Maxwellian bound is instead driven by capture at low $u$ where $(\alpha^{\rm NT})^2_{\rm max}(u)$ is very small, and  the relaxation factor is again large.

As a consequence of the discussion above, we conclude that, with the exception of the WIMP--proton SD coupling $\alpha_{\rm SD}^{p}$ at large WIMP mass ($m_\chi\gtrsim$ 1 TeV), the sensitivity on the specific choice of the velocity distribution $f(u)$ of the bounds on all the long--range couplings is of the same order of that of the corresponding couplings in the contact interaction case. 

\section{Conclusions}
\label{sec:conclusions}

In the present paper we have updated the bounds on the couplings $\alpha_{SI}^{p,n}$ and $\alpha_{SD}^{p,n}$ for a spin--independent (SI) and spin--dependent (SD) WIMP--nucleon interaction Hamiltonian in the case of a long--range interaction (i.e., for a massless mediator in the propagator). In order to do so we have used a combination of capture in the Sun (fixing the WIMP annihilation channel to $b\bar{b}$) and direct detection, assuming a standard Maxwellian $f_{\rm MB}$ for the WIMP velocity distribution (SHM). In particular, in the case of a massless mediator the usual expression used to calculate the capture rate in the Sun diverges, due to the contribution of WIMPs locked into orbits of very large semi--major axis $r_0$ by scattering events with arbitrarily low values of the momentum transfer, corresponding to low incoming WIMP speeds $u$ in the solar rest frame. Following previous analyses in the literature, we have assumed that the gravitational disturbances far from the Sun put an upper cut on $r_0$, and regularized the rate identifying $r_0$ to the Sun--Jupiter distance. We have also discussed the dependence of the SHM bounds from capture in the Sun on the Jupiter cut, and on small alterations of $f_{\rm MB}$ for $u\rightarrow$ 0, showing that they can be large for $m_\chi\gtrsim$ 1 TeV, but are moderate for smaller WIMP masses. 
Moreover, we have discussed the sensitivity of the bounds on a generic WIMP velocity distribution by obtaining halo--independent constraints using the single--stream method~\cite{Halo-independent_Ferrer2015}.

As far as the SHM bounds are concerned, our update of the combined bounds from DD and capture in the Sun on $\alpha_{SI}^{p,n}$ and $\alpha_{SD}^{p,n}$ have improved between about two and three orders of magnitude compared to the analysis of Refs.~\cite{Guo:2013ypa,Liang:2013dsa}.
In particular, compared to a contact interaction, a massless mediator systematically enhances the signal in NTs compared to that in DD. Our results remain qualitatively similar to those of Refs.\cite{Guo:2013ypa,Liang:2013dsa}: in the case of the SI couplings $\alpha_{SI}^{p,n}$ and of the SD coupling $\alpha_{SD}^{n}$ the bounds are driven by DD, while in the case of $\alpha_{SD}^{p}$ the most stringent constraint comes from capture in the Sun. 

The halo--independent bounds do not depend on the Jupiter cut needed to regularize the calculation of the latter in the Maxwellian case. We find that, with the exception of $\alpha_{\rm SD}^p$ at large $m_\chi$, the maximal weakening of each bound when a halo model different from the SHM is assumed (relaxation factor) is of the same order of that for contact interactions. This is due to the fact that for $\alpha_{\rm SI}^{p,n}$ and $\alpha_{\rm SD}^{n}$ the velocity intervals which determine both the Maxwellian and the halo--independent rates are far from the $u\rightarrow$ 0 regime responsible for the sensitivity of the expected capture rate from a massless propagator, so that the ensuing relaxation factor behaviour is similar to that in the case of a contact interaction~\cite{halo_independent_sogang_2023}: relatively moderate in the low and high WIMP mass regimes and as large as $\sim 10^2$ for $m_\chi\simeq$ 20 GeV.
On the other hand for $\alpha_{\rm SD}^p$  the exact determination of the relaxation factor becomes not reliable for $m_\chi\gtrsim$ 1 TeV, because in this case the SHM bound, driven by capture in the Sun, is very sensitive to the specific behaviour of $f_{\rm MB}$ for $u\rightarrow$ 0. 

We conclude by observing that within a simplified model where the DM particle is coupled to a massless dark photon $\gamma_D$, assuming as in Section~\ref{sec:capture} that $\chi\chi\rightarrow b\bar{b}$ drives WIMP annihilations (and that the contribution from $\chi\chi\rightarrow \gamma_D\gamma_D$ is subdominant), one has~\cite{McDermott:2010pa}:

\begin{equation}
    <\sigma v> \simeq\frac{\pi (\alpha^{p,n}_{SI})^2}{3m_\chi^2}\simeq <\sigma v>_{\rm thermal} \left(\frac{\alpha^{p,n}_{SI}}{4.2\times 10^{-3}} \right)^2\left(\frac{100~\mbox{GeV}}{m_\chi} \right)^2,
\end{equation}
\noindent with $<\sigma v>_{\rm thermal}\simeq 2.2\times 10^{-26}$ cm$^3$s$^{-1}$ the reference value that corresponds to the observed DM density for a thermal relic. This implies that, at face value, the Maxwellian bounds on $\alpha^{p,n}_{SI}$ in Fig.~\ref{fig:coupling_mx_SHM} are strong enough to overclose the Universe, and such tension is only slightly relaxed by the weaker halo--independent constraints of Fig.~\ref{fig:coupling_mx_HI}. Anyway, our phenomenological analysis is agnostic both on the ultraviolet completion of the WIMP--nucleon effective Hamiltonian of Eq.~(\ref{eq:H_long}) and on the specific mechanism that produced the DM relic abundance in the Early Universe.

\section*{Acknowledgements}
This research was supported by the National
Research Foundation of Korea (NRF) funded by the Ministry of Education
through the Center for Quantum Space Time (CQUeST) with grant number
2020R1A6A1A03047877, by the Ministry of Science and ICT with grant
number RS-2023-00241757,
and by the fund from the Institute for Basic Science (IBS) under project code IBS-R016-Y2. 
A.K. acknowledges the hospitality of the Institut d’Astrophysique de Paris (IAP).

\appendix

\section{Implementation of the experimental bounds}
\label{app:experiments}

\subsection{LZ and XENON1T and XENONnT}
\label{app:lz}
LUX--ZEPLIN(LZ) has an exposure of 3.3$\times$10$^5$ kg days. We use the efficiency provided in Fig.~2 of~\cite{LZ_2022}.
To calculate our bounds we assume 3.4 residual candidate events in the nuclear recoil energy range 1.25 keV $\le E_R \le$ 80 keV~\cite{LZ_2022}, which reproduce the published exclusion plots for a standard SI interaction.

For XENON1T we assume 7 events in the nuclear recoil energy range 1.8 keV $\le E_R \le$ 62 keV~\cite{xenon_2018} and the efficiency provided in Fig.~1 of~\cite{xenon_2018} with an exposure of 362,400 kg days.
The provided efficiencies are directly expressed in keV including the effects of quenching and energy resolution for both experiments.

In terms of XENONnT, we assume 3 events in the region of interest from 5 PE (photoelectrons) to 88 PE according to Fig. 3 of ~\cite{xenon_nT} with an exposure of 397,850 kg days. 
The efficiency is provided in Fig. 2 of ~\cite{xenon_nT} while we followed previous analyses for quenching (\cite{xenon_quenching}) and resolution (\cite{xenon_resolution}).


\subsection{PICO--60 ($C_3F_8$)}
\label{app:pico60_c3f8}
PICO--60 is a bubble chamber that detects a signal only above some value $E_{\rm th}$ of the deposited energy.
In this case the expected number of events is given by:

\begin{equation}
R=N_T MT\int_0^{\infty} P(E_R) \frac{dR}{dE_R} dE_R,
\label{eq:r_threshold}
\end{equation}

\noindent with $P(E_R)$ the nucleation probability.

For the $C_3F_8$ target material we used the total exposure~\cite{pico60_2019}, consisting in 1404 kg days at the threshold $E_{\rm th}$=2.45 keV (with 3 observed candidate events and 1 event from the expected background, implying a 90\%C.L. upper bound of 6.42 events~\cite{feldman_cousin}) and 1167 kg days at the threshold $E_{\rm th}$=3.3 keV (with zero candidate events and negligible expected background, implying an upper bound of 2.3 events at 90\% C.L.).
We have assumed the nucleation probabilities in Fig. 3 of \cite{pico60_2019} for the two runs. 

\subsection{PICO--60 ($CF_3I$)}
\label{app:pico60cf3i}

For the PICO--60 run employing a $CF_3I$ target we adopt an energy threshold of 13.6 keV and a 1335 kg days exposure. The nucleation probabilities for each target element are taken from Fig.4 in~\cite{pico60_2015}.

\subsection{Neutrino Telescopes}
\label{app:NT}

\begin{figure*}[ht!]
\centering
\includegraphics[width=7.49cm,height=6.2cm]{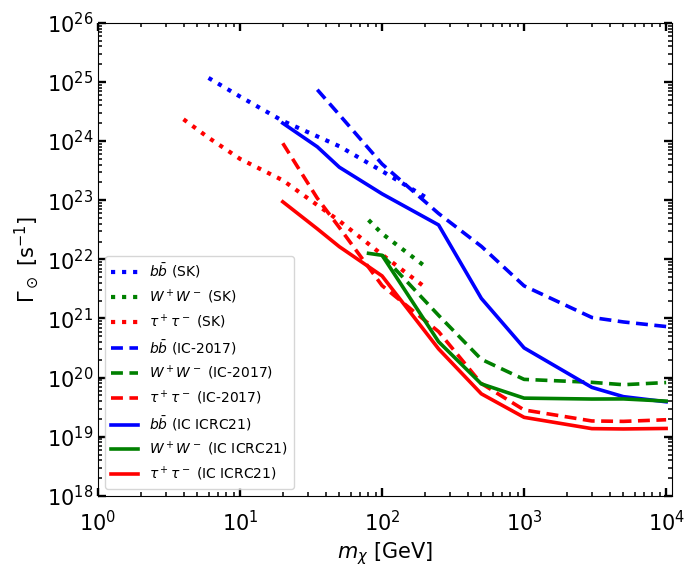}
\caption{Experimental upper bounds on $\Gamma_\odot$ (for different values of $m_\chi$) 
from the observations of Super--Kamiokande and IceCube, as discussed above. 
Bounds are shown for different WIMP annihilation channels.}
\label{fig:NT_limits}
\end{figure*}

\begin{figure*}[ht!]
\centering
\includegraphics[width=7.49cm,height=6.2cm]{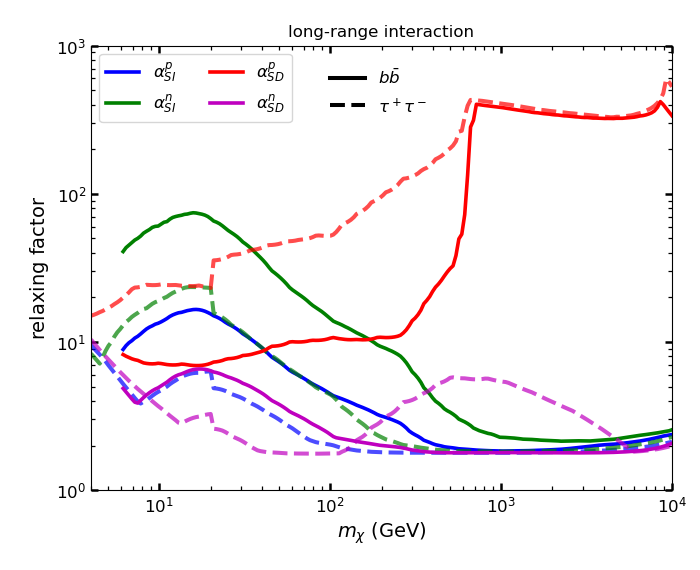}
\caption{The same as in Fig.~\ref{fig:relaxing_factor} including the $\tau\bar{\tau}$ annihilation channel.}
\label{fig:relaxing_factor_diff_channel}
\end{figure*}

Neutrino telescopes provide constraints on the neutrino flux originating from the 
annihilation of WIMPs captured in the Sun. In this work we have used the 
observations of two existing neutrino telescopes, IceCube and Super--Kamiokande, 
and the future projections of the upcoming neutrino telescope such as the Hyper--Kamiokande. 
For IceCube we have used Refs. \cite{IceCube:2016} and \cite{IceCube:2022wxw}, 
while for Super--Kamiokande we have used Ref. \cite{SuperK_2015}. 
The combined limits of Super--K and the former IceCube observations are shown by 
the dashed blue lines in Figs. \ref{fig:coupling_mx_SHM} and 
\ref{fig:coupling_mx_HI} for the SHM. 
On the other hand the combined limits 
of Super--K and the latter IceCube observations are shown by the solid blue lines.
For the halo--independent bounds (in Figs. \ref{fig:coupling_mx_HI} and \ref{fig:copulingMax_u}) we have used the combination 
of Super--K and the latter IceCube observations.
In Ref. \cite{IceCube:2016} analysing the neutrino data taken from the 
direction of the Sun for a lifetime of 532 days 
the IceCube collaboration has provided 90\% C.L. upper bounds on the WIMP annihilation rate 
$\Gamma_\odot$ for different annihilation channels ($b\bar{b}$, $W^+W^-$ and $\tau^+\tau^-$). 
For the $b\bar{b}$ channel such bound is 
$\Gamma_\odot\lesssim [7.4\times10^{24} \rm s^{-1}, 7.3\times10^{20} \rm s^{-1}]$ 
for $m_{\chi}$ in the range 35 GeV -- 10 TeV. 
Considering the new IceCube results \cite{IceCube:2022wxw}, 
the corresponding bound is provided on the SD WIMP--proton cross section (assuming the SHM). 
This bound, when converted into $\Gamma_\odot$ 
(assuming equilibrium between capture and annihilation), is found to be 
$\Gamma_\odot\lesssim [2.2\times10^{24} \rm s^{-1}, 3.9\times10^{19} \rm s^{-1}]$ 
for $m_{\chi}$ in the range 20 GeV -- 10 TeV, for the $b\bar{b}$ channel. 
The bounds from the Super--Kamiokande collaboration~\cite{SuperK_2015} are obtained 
using the data for an exposure of 3903 days. Such Super--Kamiokande bounds, 
which are expressed in terms of 95\% C.L. upper limits on the WIMP--nucleon cross section 
for $m_{\chi}$ in the range 6 -- 200 GeV, correspond to $\Gamma_\odot \lesssim [1.2\times10^{25} \rm s^{-1}, 1.2\times10^{23} \rm s^{-1}] $ 
for the $b\bar{b}$ channel. 
The Hyper--Kamiokande projections on the SD WIMP--proton cross section 
(for the $b\bar{b}$ channel) are 
taken from \cite{Bell:2021esh} (considering the uncertainty in the systematics) and are converted into the projections on $\Gamma_\odot$. These are then used to find the projections on 
different WIMP--nucleon couplings shown in Fig. \ref{fig:coupling_mx_SHM} (the purple bands). 
In Fig.~\ref{fig:NT_limits} we summarize the present experimental bounds on the annihilation rate $\Gamma_\odot$ for several annihilation channels.
In our work, with the goal to obtain conservative bounds, we consider only annihilations to $b\bar{b}$, which, among the different channels, provides the less constraining bound\footnote{Note that, 
annihilation to light quarks, which are stopped in the solar plasma before hadronizing and decaying, have also been considered in the literature, but are particularly challenging to the NT's~\cite{capture_light_quarks1,capture_light_quarks2,capture_light_quarks3, NT_DD_Blennow2015}.}. For completeness, in Fig.~\ref{fig:relaxing_factor_diff_channel} we provide a plot of the relaxation factor of Fig.~\ref{fig:relaxing_factor} also for the $\tau\bar{\tau}$ annihilation channel, and compare the results to the $b\bar{b}$ case. 

\end{document}